\documentclass{article}[12pt]
\usepackage{amsmath,amssymb}
\usepackage{graphicx}%
\usepackage{verbatim}

\newcommand{\TwoFig}[4]{%
\begin{tabular}{lr}
\parbox{8cm}{\includegraphics[width=7cm]{#1}}  & \parbox{8cm}{\includegraphics[width=7cm]{#2}} \\
\parbox{8cm}{\vspace{7pt}\refstepcounter{figure}Figure \thefigure.\quad #3\vfill} & \parbox{8cm}{\vspace{7pt}\refstepcounter{figure}Figure \thefigure.\quad #4\vfill} \\
\end{tabular}
}

\newcommand{\Fig}[3]{%
\begin{center}
\parbox{#2cm}{%
\refstepcounter{figure}\includegraphics[width=#2cm]{#1} \noindent Figure \thefigure:\quad
#3}\end{center}}
\begin{document}

\begin{center}
{\bf \Large Complete cosmological evolution model of a classical scalar field with a Higgs potential. III. Features of phase trajectory flows\\[12pt]
Yu. G. Ignat'ev, A. R. Samigullina }\\
Physics Institute of Kazan Federal University,\\
Kremleovskaya str., 35, Kazan, 420008, Russia.
\end{center}

\begin{abstract}

The flows of phase trajectories of cosmological models based on the vacuum classical Higgs field and their behavior on the Einstein-Higgs surface near singular points of a dynamical system are investigated by numerical simulation. The influence of the singular points on the process of phase flow splitting is shown and the efficiency of the Einstein-Higgs hypersurface application is demonstrated.\\

{\bf keywords}:cosmological models, Higgs fields, Einstein-Higgs hypersurface, global behavior.

\end{abstract}

\section{Dynamical system cosmological models}
We have already formulated and partially investigated the complete cosmological model \cite{IgnI} based on the classical scalar Higgs field. In particular, we have introduced the concept of the Einstein-Higgs hypersurface important to understand the behavior of the cosmological models. This model was generalized to the asymmetric scalar Higgs doublet containing classical and phantom scalar fields \cite{IgnII}  and investigated by the methods of qualitative and numerical analysis. In our previous works based on the model of cosmological evolution of the classical scalar Higgs field, the detailed qualitative and numerical analysis of the behavior of the corresponding cosmological models was performed. In the present work, we study the flow of phase trajectories for the considered dynamic model, because individual phase trajectories may possibly describe random features in the behavior of the cosmological models. To elucidate the general laws of the system behavior, in particular, near its singular points, it is necessary to study the phase flows in which each trajectory is generated with sufficiently close initial conditions. Recall that the cosmological model based on the classical vacuum scalar Higgs field in the phase space $R_3=\{\Phi,Z,h\}$  (the scalar potential, its derivative, and the normalized Hubble parameter) is described by the independent three-dimensional dynamical system:
\begin{equation} \label{Eq1}
\Phi'=Z,
\end{equation}	
\begin{equation} \label{Eq2}
Z'=-3hZ-e\Phi+\alpha_m\Phi^3,
\end{equation}	
\begin{equation} \label{Eq3}
h'=-3h^2+\frac{e\Phi^2}{2}-\frac{\alpha_m\Phi^4}{4}+\lambda_m ,
\end{equation}
where the derivatives are taken with respect to time and are expressed in the Compton units  $\tau=mt$; $\lambda_m=\lambda_0-m^4/4\alpha_m$  and $\alpha_m$ are the renormalized cosmological and self-action constants, respectively; and $e=\pm 1$  is the indicator.
In addition, it is necessary to consider the equation for the Einstein-Higgs hypersurface being, as a matter of fact, the Einstein equation $^4_4$ ,
and on the other hand, the first partial integral of dynamical system \eqref{Eq1}-\eqref{Eq3}:
\begin{equation} \label{Eq4}
3h^2-\mathcal{E}_m(\Phi,Z)=0,
\end{equation}
where
\begin{equation} \label{Eq5}
\mathcal{E}_m(\Phi,Z)=\frac{Z^2}{2}+\frac{e\Phi^2}{2}-\frac{\alpha_m\Phi^4}{4}+\lambda_m\geqslant 0
\end{equation}
is the effective energy of dynamical system \eqref{Eq1}-\eqref{Eq3}. We now write formulas for the physical characteristics of the cosmological model: invariant cosmological acceleration
\begin{equation} \label{Eq6}
\Omega \equiv \frac{a\ddot{a}}{\dot{a}^2}\equiv1+\frac{\dot{h}}{h}^2
\end{equation}
and invariant curvature of the Friedman space
\begin{equation} \label{Eq7}
\sigma\equiv\sqrt{R_{ijkl}R^{ijkl}}=H^2\sqrt{6(1+\Omega^2)}\equiv\sqrt{6}\sqrt{H^4+(H^2+\dot{H})^2}\geqslant 0
\end{equation}
Let us write Eq. \eqref{Eq3} in another form:
\begin{equation} \label{Eq8}
h'=-\frac{Z^2}{2} (\leqslant 0),
\end{equation}
being a differential-algebraic consequence of formulas \eqref{Eq3} and \eqref{Eq4} convenient for an analysis of the behavior of phase trajectories. According to this formula, the Hubble constant cannot increase with time in the cosmological model with the classical scalar Higgs field.
Moving on to numerical modeling of the dynamical system under study, we first note that some essential features in the behavior of this system have already been revealed based on numerical modeling \cite{IgnI}. According to that work, we define dynamical system \eqref{Eq1}-\eqref{Eq3} by two ordered lists including three dimensionless parameters $P=[e,\alpha_m,\lambda_m]$  and initial conditions $I=[\Phi_0,Z_0,\varepsilon]$ , where $\varepsilon=\pm 1.$ Here $\varepsilon=+1$  corresponds to the expansion phase at the initial time  $t_0$, whereas $\varepsilon=-1$  corresponds to the compression phase at that time. According to the foregoing, the initial value of the Hubble constant is determined by Einstein's equation \eqref{Eq4} from which we obtain
\begin{equation} \label{Eq9}
h_0=\pm\frac{1}{2}\sqrt{\frac{Z^2_0}{2}+\frac{e\Phi^2_0}{2}-\frac{\alpha_m\Phi^4_0}{4}+\lambda_m}\equiv\frac{\varepsilon}{3}\sqrt{\mathcal{E}^0_m}
\end{equation}
Dynamical system \eqref{Eq1}-\eqref{Eq3} is an autonomous system of ordinary differential equations invariant under the translation $t\rightarrow t_0+t$. Therefore, any arbitrary value $t_0$  can be chosen as the initial time for the initial conditions. We set this value equal to zero. In this case, we can also consider the dynamical system states at a negative time $t_0< 0$ .
Numerical integration by the Runge-Kutta method (dverk78) was performed using the mathematical package Maple with relative accuracy and absolute calculation accuracy equal to $10^{-7}$ . In this work, we use results of our previous works \cite{IgnIII},\cite{IgnIV}, in particular, results of our analysis of the singular system points. Below we consider two radically different cosmological models with positive and negative cosmological constant.
\section{Phase flows for the model with the positive cosmological constant $\Lambda_m = 0.1$}
\subsection{Model with the parameters P=[1, 1, 0.1]}
Figures 1-3 show the phase trajectories obtained by numerical integration of dynamical system \eqref{Eq1}-\eqref{Eq3} with the parameters P=[1, 1, 0.1], and Figs.4 and 5 show the plots of evolution of the effective energy density $E_m(t)=\mathcal{E}_m(\Phi(t),Z(t))$  and of the cosmological acceleration $\Omega(t)$. Here $Z(0)=1$  everywhere. Note that in this case, the singular points $M_{11}$, $M_{12}$, and $M_{22}$ are saddle points, the point $M_{+}$  is attractive, and the point $M_{-}$ is repelling.

\TwoFig{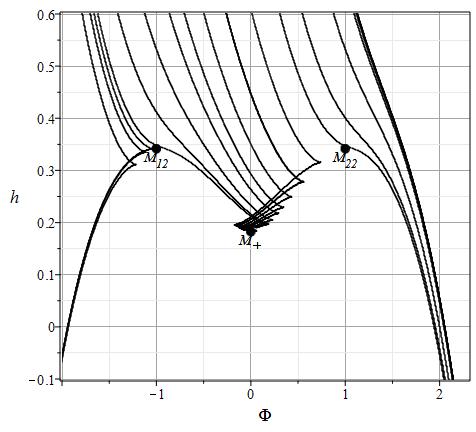}{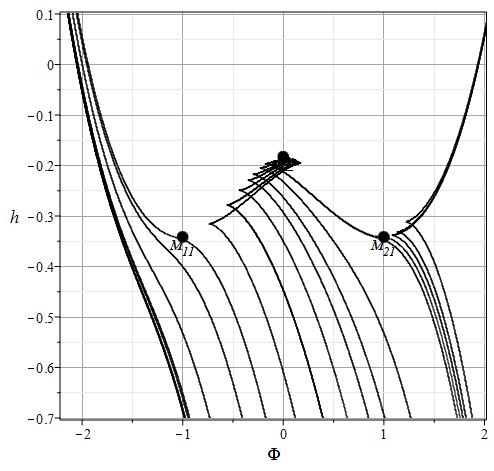}{Phase trajectories of dynamical system \eqref{Eq1}-\eqref{Eq3} with the parameters P=[1,1,0.1] in the plane $S=\{\Phi,h\}$  against the background of the map of the singular points for $h(0)=h_{+}>0$.}{Phase trajectories of dynamical system \eqref{Eq1}-\eqref{Eq3} with the parameters P=[1,1, 0.1] in the plane $S=\{\Phi,h\}$  against the background of the map of the singular points for $h(0)=h_{-}<0$.}
\TwoFig{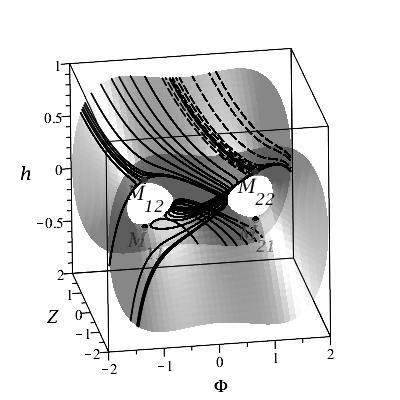}{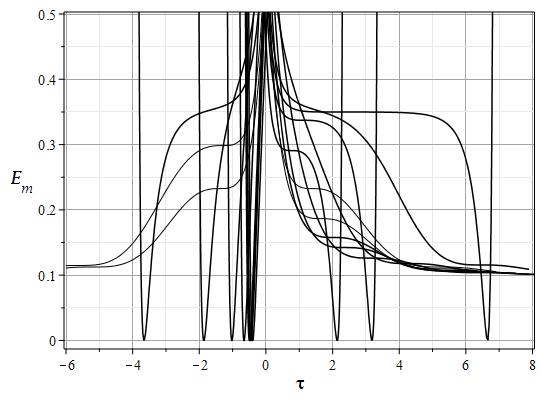}{Phase trajectories of dynamical system \eqref{Eq1}-\eqref{Eq3} with the parameters P=[1, 1, 0.1] on the Einstein-Higgs hypersurface. Here the solid curves are for the initial conditions $\Phi(0) = [-1.5, -1.47, -1.45$, $-1.425, -1.4,-1.25, -1, -0.7, -0.5,-0.25,-0.0001]$, the dashed curves are for $\Phi(0)=[0.0001, 0.25, 0.3$, $0.35$, $0.36, 0.38, 0.4, 0.45, 0.5, 0.7, 1, 1.5]$, and $h(0)=h_{+}$  everywhere.}{Evolution of the effective energy Em with the initial conditions $\Phi(0) = [-1.5,-1.47, -1.45,-1.425$,$-1.4,-1.25,-1,-0.7$, $-0.5$, $-0.25,-0.0001]$ (from left to right), $h(0)<0$ (on the left), and $h(0)>0$ (on the right).}
\Fig{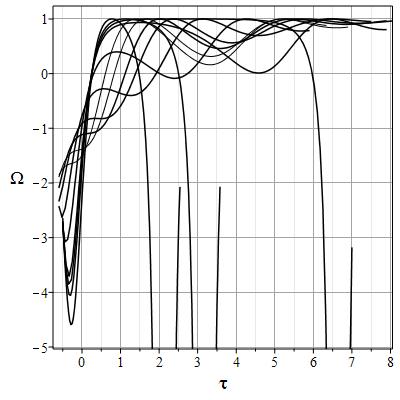}{8}{Evolution of the cosmological acceleration $\Omega$  with the initial conditions $\Phi(0) = [-1.5,-1.47,-1.45,-1.425,-1.4,-1.25,-1,-0.7,-0.5,-0.25,-0.0001]$ (from left to right).}
\subsection{Model with the parameters $P=[-1,-1, 0.1]$}
Figures 6 and 7 show the phase trajectories obtained by numerical integration of dynamical system \eqref{Eq1}-\eqref{Eq3} with the parameters $P=[-1,-1, 0.1]$, and Figs. 8 and 9 show the plots of the evolution of the effective energy density $E_m(t)=\mathcal{E}_m(\Phi(t),Z(t))$  and of the cosmological acceleration $\Omega(t)$. Here $Z(0)=1$  everywhere. In this case, there are only two singular points: the attractive point $M_{+}$  and the repelling point $M_{-}$ .

\section{Phase flows for the model with negative cosmological constant $\Lambda_m=-0.1$}
\subsection{Model with the parameters $P=[1, 1,-0.1]$}
Figures 10 and 11 show the phase trajectories obtained by numerical integration of dynamical system \eqref{Eq1}-\eqref{Eq3} with the parameters $P=[1, 1,-0.1]$, and Figs. 12 and 13 show the evolution of the effective energy density  $E_m(t)=\mathcal{E}_m(\Phi(t),Z(t))$ and of the cosmological acceleration $\Omega(t)$ . Here $Z(0)=1$  everywhere. In this case, there are four saddle singular points $M_{11}$, $M_{12}$, $M_{21}$ and $M_{22}$ .
\TwoFig{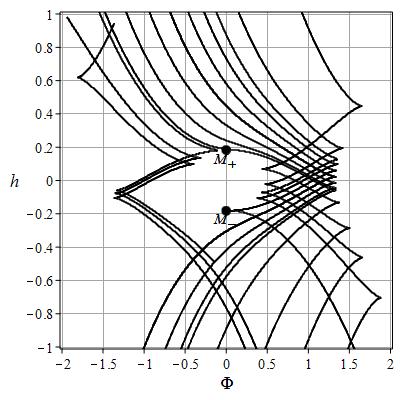}{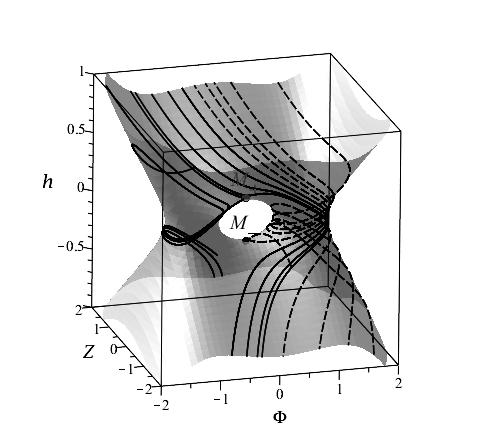}{Phase trajectories of dynamical system \eqref{Eq1}-\eqref{Eq3} with the parameters P=[1,1,0.1] in the plane $S=\{\Phi,h\}$  against the background of the map of the singular points for $h(0)=h_{+}>0$.}{Phase trajectories of dynamical system \eqref{Eq1}-\eqref{Eq3} with the parameters $P = [-1,-1, 0.1]$ on the Einstein-Higgs hypersurface with the initial conditions $\Phi(0)=[-1.5,-1, -0.7,-0.5,-0.25,-0.0001]$ (the solid curves) and $\Phi(0) = [0.0001, 0.25, 0.5, 0.7, 1, 1.5]$ (the dashed curves).}
\TwoFig{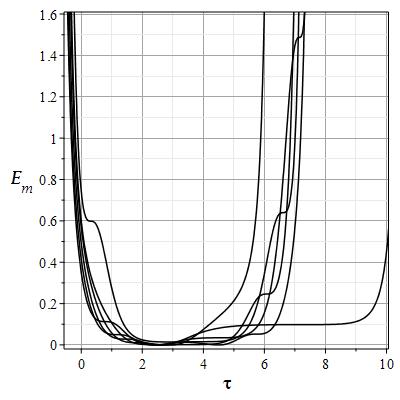}{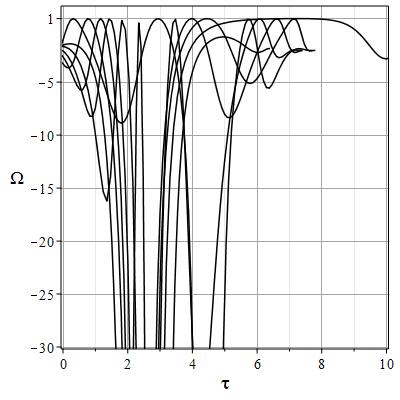}{Evolution of the effective energy $E_m$  with the parameters $P=[-1,-1, 0.1]$ and the initial conditions $\Phi(0)=[0.0001,0.25,0.5,0.7,1,1.5]$. Here $Z(0)=1$  and  $h(0)=h_{+}$  everywhere.}{ Evolution of the cosmological acceleration $\Omega$  with the parameters  $P = [-1,-1, 0.1]$ and the initial conditions $\Phi(0)= [0.0001, 0.25, 0.5, 0.7, 1, 1.5]$. Here $Z(0)=1$  and $h(0)=h_{+}$ everywhere.}
\TwoFig{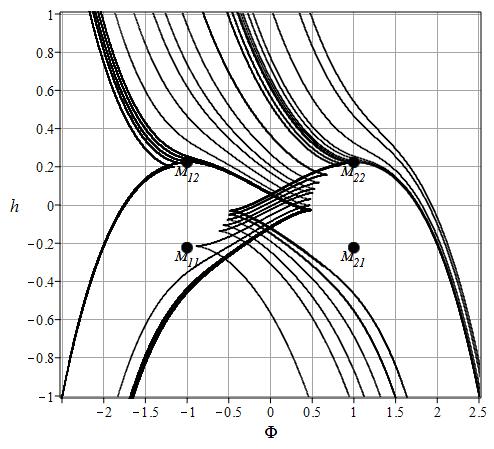}{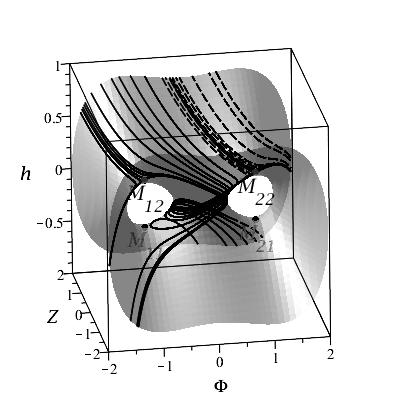}{Phase trajectories of dynamical system \eqref{Eq1}-\eqref{Eq3} with the parameters $P=[1, 1,-0.1]$ in the plane $S=\{\Phi,h\}$  against the map of the singular points for $h(0)=h_{+}>0$.}{ Phase trajectories of dynamical system \eqref{Eq1}-\eqref{Eq3} with the parameters $P = [1, 1, -0.1]$ on the Einstein-Higgs hypersurface with the initial conditions   $\Phi(0)= [-1.5, -1.47,-1.45,-1.425,-1.4,-1.25,-1$, $-0.7,-0.5,-0.25,-0.0001]$ (the solid curves) and  $\Phi(0) = [0.0001, 0.25, 0.3, 0.35, 0.36, 0.38, 0.4$, $0.45, 0.5, 0.7, 1, 1.5]$ (the dashed curves).}
\TwoFig{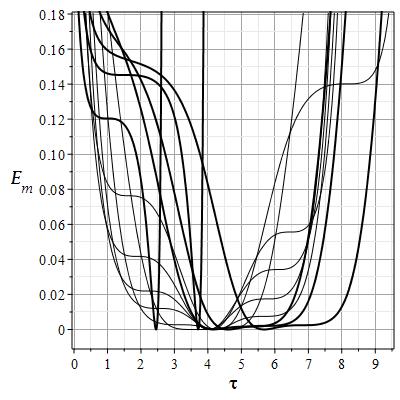}{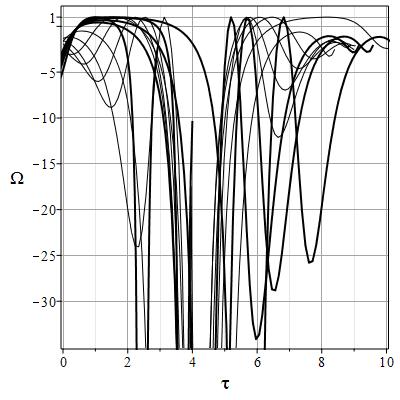}{Evolution of the effective energy $E_m$  with the parameters $P=[1,1,-0.1]$ and the initial conditions  $\Phi(0)=[-1.5,-1.47,-1.45,-1.425,-1.4,-1.25,-1$, $-0.7,-0.5,-0.25,-0.0001]$. Here $Z(0)=1$  and $h(0)=h_{+}$  everywhere.}{Evolution of the cosmological acceleration $\Omega$  with the parameters $P=[1,1,-0.1]$ and the initial conditions  $\Phi(0)=[-1.5,-1.47,-1.45,-1.425,-1.4,-1.25,-1$, $-0.7,-0.5,-0.25, -0.0001]$. Here $Z(0)=1$   and $h(0)=h_{+}$  everywhere.}
\section{Analysis of the results}
As can be seen from the phase portraits of the dynamical system in the $\{\Phi,H\}$  plane shown in Figs. 1, 2, 6, and 10, the phase flow for dynamical system \eqref{Eq1}-\eqref{Eq3} splits into two flows near the singular points of the system. The mechanism of flow splitting is similar to the action of underwater stones in the fast flow of the river in which the role of stones is played by the singular points attracting the phase trajectories (the attracting singular points) or repelling them (the saddle singular points). In this case, the analogy with the water flow is very close considering that the phase trajectories can only roll down $(h'\leqslant0)$, so that the variable $h$  in the model with the classical Higgs field plays the role of the water level in this analogy. The three-dimensional phase portraits of the dynamical systems in Figs. 3, 7, and 10 vividly demonstrate rolling down of the phase flows of the dynamical system.	
The results of numerical modeling presented above demonstrate once again the efficiency of analysis of the cosmological model based on the study of the geometry of the Einstein-Higgs hypersurface playing the role of the geometry of the surface relief which can be used to predict easily the global behavior of the cosmological model.

This paper has been supported by the Kazan Federal University Strategic Academic Leadership Program (PRIORITY-2030).


\end{document}